\documentclass[12pt,preprint]{aastex}
\shorttitle{Solar Gravitational Deflection}
\shortauthors{Fomalont et al.}
\begin{document}

\title{Progress in Measurements of the Gravitational Bending of Radio Waves Using
the VLBA}

\author{E.~Fomalont}
\affil{National Radio Astronomy Observatory, Charlottesville, VA 22903}
\email{efomalon@nrao.edu}

\author{S.~Kopeikin}
\affil{Department of Physics \& Astronomy, University of Missouri, Columbia, MO 65211}
\email{kopeikins@missouri.edu}

\author{G.~Lanyi}
\affil{Jet Propulsion Laboratory, California Institute of Technology, Pasadena, CA 91109}
\email{gelanyi@jpl.nasa.gov}

\author{J.~Benson}
\affil{National Radio Astronomy Observatory, Socorro, NM 87801}
\email{jbenson@nrao.edu}

\begin{abstract} 

We have used the Very Long Baseline Array (VLBA) at 43, 23 and 15 GHz
to measure the solar gravitational deflection of radio waves among
four radio sources during an 18-day period in October 2005.  Using
phase-referenced radio interferometry to fit the measured phase delay
to the propagation equation of the parameterized post-Newtonian (PPN)
formalism, we have determined the deflection parameter $\gamma =
0.9998\pm 0.0003$ (68\% confidence level), in agreement with General
Relativity.  The results come mainly from 43 GHz observations where
the refraction effects of the solar corona were negligible beyond 3
degrees from the sun.  The purpose of this experiment is three-fold:
to improve on the previous results in the gravitational bending
experiments near the solar limb; to examine and evaluate the accuracy
limits of terrestrial VLBI techniques; and to determine the prospects
and outcomes of future experiments.  Our conclusion is that a series
of improved designed experiments with the VLBA could increase the
presented accuracy by at least a factor of 4.

\end{abstract}


\section{Introduction}

The parameterized post-Newtonian (PPN) formalism was designed by
Nordtvedt and Will \citep{wil93} to analyze plausible deviations in
the general theory of relativity in the case of a
spherically-symmetric gravitational field of a massive body.  The PPN
formalism introduces several phenomenological parameters that are
targets for experimental measurements, and the propagation of
electromagnetic waves is affected strongly by one of them, $\gamma$.
This parameter empirically measures the deviation from the linear
approximation of General Relativity in which case the value of $\gamma
=1.0$ \citep{ein16}: The Newtonian limit in this linear approximation
would imply a value $\gamma=0$ (see equation 7.22 in \cite{wil93}).
The parameter $\gamma$ can be determined precisely and unambiguously
by measuring the deflection of electro-magnetic radiation by the
gravitational field of the sun, with the GR value of $1.750''$ at the
solar limb.  Such experiments have been performed since 1919
\citep{dys20}, and the most accurate interferometric observations have
shown that $\gamma$ is consistent with unity to an accuracy of less
than one part in 1000 \citep{sha64,fom75,rob91,leb95,sha04}.
Recently, \cite{ber03} have made use of the Cassini tracking data in
2002 to set the upper limit of $\gamma$ to a few parts in
100,000. Somewhat less accuracy has been obtained with another analysis
of the data \citep{and04}, and there are some concerns about the effect
of the motion of the solar barycenter \citep{kop07}. Hence, additional
accurate and independent measurements are important to determine the
value of the fundamental parameter, $\gamma$, using the
radio-interferometric and other techniques.

With the demonstrated accuracy of the VLBA to measure relative
positions of radio sources to $0.01$ mas \citep{fom03,bru06}, the
gravitational bending could be measured potentially with the
radio-interferometric technique to a few parts in 100,000, although
the coronal refraction when observing within a few degrees of the sun
produces large path-length changes.  In this paper, we present the
results from the measurement of $\gamma$ performed with the Very Long
Baseline Array (VLBA) in October 2005, and we suggest how more
accurate measurements of $\gamma$ can be obtained.

\section{The Observations and Data Reduction}

\subsection{The Observational Strategy and Parameters}

Every year in early October, the sun passes in front of the strong
radio source 3C279.  By measuring the change of its angular position,
caused by the gravitational bending effect, with respect to other
nearby sources over a period of several weeks, the gravitational
parameter $\gamma$ can be accurately determined, and we used the
opportunity in 2005 to conduct a dedicated VLBA experiment.  In
contrast to several previous interferometric experiments that measured
the relative position of 3C279 with respect to 3C273 using the group
delay \citep[e.g.][]{leb95}, our observations used the phase delay, a
more accurate astrometric quantity, to measure the change of relative
position of 3C279 with respect to that of several fainter sources
within a few degrees in the sky.

For phase referencing observations among close sources, the choice of
sources to observe must be balanced between two competing factors: the
closer the radio sources are in the sky, the more accurate their
relative positions can be determined \citep{fom05}; the farther apart
the radio sources are in the sky, the greater the relative
gravitational deflection among them.  The angular scale pertinent to
this balance is related to the closest that successful source
observations can be made near the sun, about two degrees.  The
sensitivity of the VLBA determines whether there are any sources
within a few degrees of 3C279 that can be detected in order to measure
their accurate positions.  From an existing catalog of compact and
relatively bright sources \citep{pet06}, several candidates were
found, and after brief VLBA observations, three were chosen for this
experiment, and their configuration in the sky is shown in Fig.~1.
The apriori positions and total flux densities of the sources are
shown in Table 1, and their positions, estimated uncertainties and
compactness have been determined within the International Celestial
Reference Frame system to an accuracy of $<1$ mas.  Even the weakest
of the sources can be detected by the VLBA within about 20 seconds of
integration time.

The sources were observed with the VLBA on eight days (sessions)
centered around the October 8 solar occultation of 3C279: October 1,
5, 6, 7, 9, 10, 11 and 18, for a period of six hours each session when the
sources were above $10^\circ$ elevation at most of the VLBA antennas.
Because the ionized plasma in the solar corona produces a
frequency-dependent deflection that increases inversely with
wavelength squared, observations at the highest routinely available
VLBA frequency of 43 GHz were chosen to minimize the coronal
deflection.  We also observed the sources at 23 GHz and 15 GHz in
order to estimate the coronal deflection that was present.

Since the VLBA cannot observe simultaneously among 15, 23 or 43 GHz,
it is not possible to remove the coronal refraction
instantaneously\footnote{The 2.3/8.4 GHz simultaneous frequency
system, which is available on the VLBA and used for most astrometric
observations to remove the ionosphere refraction \citep{ma98}, is
subject to severe scattering by the coronal refraction within about
four degrees of the sun. Previous observations to measure the solar
deflection at these low frequencies alone did not produce $\gamma$
accuracies better than the $10^{-3}$ level.  Hence, higher frequencies
were used \citep{leb95}.}.  Thus, we had to choose a switching cycle
among the sources and frequencies which minimized the effects of phase
fluctuations from the troposphere and corona.  Because the short-term
tropospheric phase fluctuations at 43 GHz can limit the coherence
time-scale to less than a minute \citep{bea95}, it is crucial to
switch observations (scans) among the sources within this period in
order to keep the phase coherent between source scans.  Since a change
in the observing frequency at the VLBA takes at least 15 sec, any
frequency switching would have considerably lengthened the time
between scans, beyond the nominal coherence time.  Thus, we decided
switch among the sources at one frequency with 40-sec scans for about
20 minutes.  Then, change frequencies and switch among the sources for
another 20 minutes.  This observing scheme is illustrated in Fig.~2.
Essentially, each 20-minute period of source switching can determine
their accurate relative positions at one frequency.  The slower
cycling among the frequencies could determine the different positions
among the frequencies caused by the coronal refraction, albeit on a
relatively slow time scale of about one hour.

The relevant observational parameters were as follows \citep{wal95}.
At each of the three observing bands, we observed at four frequencies
(IF's), each with a bandwidth of 16 MHz: 14.93, 14.98, 15.20, 15.37
GHz; 23.02, 23.07, 23.29, 23.46 GHz; 42.73, 42.78, 43.00, 43.17 GHz.
Each IF was separated into 16 frequency channels, each of width 1 MHz,
and then sampled with two bits.  The separation of the frequencies at
each band permitted the determination of the group delay (phase slope
versus frequency) to improve the tropospheric delay estimates using
many calibrator observations at the beginning and end of each day.  We
also included several observations of the strong sources 3C273 and
J1310+334 over the day to monitor the instrumental delay changes.  The
observing schedule for the eight sessions was identical apart from
slightly different start and stop times.
 
\subsection {Initial Data Calibration, Editing and Averaging}

The data from all antenna-pairs were processed using the VLBA
correlator in Socorro, NM, and the output visibility data were
averaged into 2-sec samples.  The apriori model in the correlator uses
the NASA CALC software package (http://www/gemini.gsfc.nasa.gov/solve)
that includes the gravitational bending, assuming $\gamma=1$.  Hence,
the fitting parameter used in the data analysis is the departure of
$\gamma$ from unity.

The standard VLBA calibrations and editing were made on the correlated
data \citep{ulv01} independently for 15 GHz, 23 GHz and 43 GHz: 1) The
apriori amplitude calibrations of each antenna, from the monitored
system temperature and aperture efficiency, as well as the correlator
bias, were applied; 2) Data were flagged during receiver/antenna
malfunctions, antenna mis-pointing, and when the source elevation for
any antenna was below $10^\circ$.  3) Improved parameters for the
earth-orientation and rotation, polar motion, antenna motions and
nutation were available from on-line services and applied to the
correlated data; 4) The calibrator observations at the beginning and
end of each session were analyzed to determined corrections to the
zenith path delay used in the correlator model \cite{mio05}.

The data from the correlator for each observing frequency contained 64
streams: four IF frequencies, each split into 16 frequency channels.
The relative gains and phases among the streams change slowly and were
determined using several observations in each session of strong
sources, 3C273 and J1310+332 (and 3C279 when not too close to the
sun).  The relative gain and phase determination among the frequency
streams varied about 5\% and $15^\circ$, respectively, over a session,
so the changes were easily corrected.  The 64 data streams were then
coherently averaged to form one data stream with high signal to noise
ratio.  These data at 15 GHz, 23 GHz and 43 GHz were then used to
determine the shorter time-scale phase changes induced by the
troposphere and solar corona, described in the next section.

\subsection {Tropospheric Editing}

When observing at high radio frequencies, small \lq\lq clouds\rq\rq~of
water vapor pass through each antenna beam and produce variable delays
of a few millimeters (about $100^\circ$ at 43 GHz) over time scales of
a minute or less.  These induced phase variations limit the
astrometric and imaging accuracies.  An example of relatively stable
phase behavior is shown in Fig.~3a.  Since 3C279 is a strong source,
its phase can be determined for each 2-sec sampled data point over a
40-sec scan (closely-spaced points), and over this period the typical
phase variation over a scan is about a $30^\circ$.  In order to have
{\it coherent} observations, the phase between the consecutive 3C279
scans must be unambiguously connected with no lobe ambiguities,
otherwise the residual phase of the other sources(their measured phase
minus the interpolated phase from 3C279) will be grossly in error.
For the BR and MK baselines, the phase connection between scans of
3C279 is obvious, although residual phase offsets of $50^\circ$ do
occur.  However, for the SC baseline, the scan to scan phase changes
of 3C279 are larger and in some cases (between 18:41 to 18:42) it is
not clear whether the phase increased or decreased between scans, so
this period with SC is of dubious astrometric quality.

Using plots similar to that of Fig~3a for all observations and
frequencies, we edited periods when the phase coherence was poor.
These periods often were associated with inclement weather conditions
at an antenna site.  The percentage of data removed was 4\% at 15 GHz,
7\% at 23 GHz and 17\% at 43 GHz, mostly for the Saint Croix VI and
Hancock, NH antennas that are at relatively humid locations.

\subsection {Coronal Editing}

In Fig.~3b we show a 2-min segment of data at 23 GHz for 3C279 on
October 9 when the source was about $1.7^\circ$ from the sun.
Although the phases are coherent over 2 seconds, there are rapid
changes of phase over 10 seconds or longer, hence the phase cannot be
connected between scans.  Notice that the phase changes for the
shortest baseline (bottom plot) of 240 km are as variable as
those for the longer baselines.  This indicates that the coronal
refraction significantly varies over second time-scales with a linear
scale-size smaller than a few hundred kilometers in the solar
corona\footnote{the analysis of the statistical properties of the
phase data when the sources were close to the sun can provide
information on the velocity, density and size of the coronal turbulent
cells.  This will be reported elsewhere.}.

We found that when a source was closer than $\sim 3^\circ$ from the
sun, phase-stable observations often were not possible because of the
severe and short term coronal refraction at all frequencies.  Thus,
all observations on October 6 and 7, when three of the four sources
were within this angle from the sun, were astrometrically worthless.
Other data not used because of coronal turbulence were for J1248 and
3C279 on October 5 and 3C279 on October 9.

\section{Astrometric Analysis} 

\subsection {The Removal of the Source Structure}

Although all of the sources are dominated by a compact core component,
extended emission is also present and often associated with a jet
component that emanates from the core.  This structure causes two
astrometric problems: the peak brightness of the core (best definition
of the location of the source) is resolution dependent, and the
structure may vary with time.  We, thus, determined the source
structures at each frequency on October 1 and October 18 using the
self-calibration algorithm \citep{cor89}.  Since this algorithm gives
no positional information about the source, we arbitrarily placed the
maximum brightness for each source at each frequency at the assumed
apriori position of the source given in Table~1.  {\it This convention
has no effect on the astrometric results since the analysis determines
the residual true sky position of the peak of the source, relative to
the assumed model and apriori position.}

Because significant structure and intensity changes in most
extra-galactic source radio emission occur over months of time, little
change was expected over the 17-day period of this experiment
\citep{hug92}.  Changes in structure can also be indirectly inferred
also by the variability of the peak flux density of the sources that
are given in Table~2.  The only significant variability occurred for
3C279 and other
monitoring\footnote{http://www.vla.nrao.edu/astro/calib/polar/2005}
showed that the source reached a maximum of 20 Jy at 23 GHz in early
2005 and decreased through the year.  Hence, the decrease in flux
density is real and the structure did change slightly over the
experiment.  Comparison of the accurate structures for 3C279 on
October 1 and October 18 showed that minor variations occurred near
the position 0.2 mas west and 0.1 mas east of the radio core at 43
GHz, but that the apparent position change of the peak brightness at
43 GHz and 23 GHz was less than 0.02 mas.

The source structures will be discussed and displayed in more detail
elsewhere (Fomalont et al, in preparation), but a summary follows: The
sources J1246 and J1248 are nearly point sources at 43 GHz, with a
faint component at 15 GHz about 0.5 mas to the west and 1.4 mas to the
north-east, respectively.  The source J1304 is a nearly equal double at
43 GHz with a separation of 0.8 mas, and 3C279 has a jet component
which extends about 1 mas south-east of the core.

Since the source structures were essentially unchanged over the 18-day
period of the experiment, we removed their effects by dividing the
visibility data with the visibility data model associated with the
source structure (averaged from the Oct 1 and Oct 18 images) at each
frequency.  This produced a {\it structure-free} (effectively now a
point source) data set for each source and frequency.  Any resolution
differences among the observing sessions will not produce a change in
the location of the peak brightness of this revised data set.

\subsection {Phase Referencing to Obtain the Residual Phase}

After the initial calibrations, editing, and removal of the source
structure, we then used the standard phase referencing technique on
the visibility data to determine the relative position among the
sources \citep{bea95}.  This technique uses one of the sources as the
main reference to determine the phase error associated with each
antenna for each reference source scan.  This phase error is then
interpolated between each reference scan, and applied to the other
sources to obtain their residual phases.  This technique is
illustrated in Fig.~3a.  Since the switching time and proximity in the
sky among the sources were small, the phase errors associated with the
reference source are virtually the same as that for all sources.
Hence, any {\it residual} phases found for the non-calibrator sources
are predominantly associated with a position offset from its apriori
value.  We used 3C279 as the main phase reference on October 1, 11 and
18 when it was not too close to the sun, and J1304 as the main phase
reference on the other days.  The phase referencing was done
independently for each frequency.

\subsection {Source Position Determination from the Residual Phase}

Two analysis methods were used to determine the radio source positions
from the measured residual phases for each source and frequency.  The
first method used the basic interferometric Fourier imaging and
deconvolution techniques to produce an image of each source.  We then
determined the location of the peak brightness of the source (all
nearly point sources since the structure component has been removed),
and an error estimate based on the quality of the image and the
signal-to-noise ratio at the peak intensity.  For the second method,
we fit the measured residual phases directly to a point source model
and determined by a least-square analysis the position offset from the
apriori position and the error estimate based on the phase deviations
from the best fit.

Although the relative source positions at each frequency could be
derived from every 20-min segment of data, the expected change of
position during each session caused by the coronal bending and by any
small departures of $\gamma$ from unity was estimated to be less than
0.20 mas at 15 GHz, and smaller at the higher frequencies.  Hence, the
data at each of the frequencies for an entire session could be
analyzed with no significant position smearing.  However, for 3C279 on
October 10 when the coronal bending significantly changed, the
analysis was split into two half-session periods.

Both methods used the same set of data, and they provided somewhat
different positions and errors that are associated with the different
effective weighting of the data points.  For the source position, we
used the average position obtained from the two methods.  The methods
provided three position errors, one from each analysis method and one
also from the difference between the two position determinations.  We
chose the largest of the three as the error estimate.  Finally, we
adopted a floor uncertainty that was based on the theoretical
signal-to-noise ratio of the data: 0.02 mas E/W and 0.03 mas N/S for
3C279, and 0.05 ms E/W and 0.07 mas N/S for the other three sources.

The resulting source positions from the above analyses, made
independently for each frequency, are shown in Table~3.  In order to
have a one fiducial reference point over the entire experiment, we
have listed in column 1 the relative positions of 3C279, J1254 and
J1256 with respect to J1304 for all days.  This requires only a
trivial position translation to J1304 of the positions on days when
3C279 was used as the phase reference\footnote{Because the
experimental errors are dominated by antenna-based residual
tropospheric and ionospheric errors, rather than by receiver noise and
other baseline-stochastic processes, there is no additional
information by using the results from all six source-pairs, compared
with the non-degenerate set of the three source-pairs listed.}.  The
distance of the source from the sun in degrees in mid-session is shown
in column 2.  When one or more of the sources were within $2.5^\circ$
of the sun, the coronal turbulence was often too severe to obtain
phase coherence and no positions could be determined.  For this
reason, there are no October 6 and 7 entries in the table.  On October
10 when 3C279 was relatively close to the sun, the analysis was made
on the first half and second half of the session because of the
relative large change in coronal refraction on this day.

For observations on October 1 and 18, when the sources were
sufficiently far from the sun, the maximum expected relative
gravitational bending was 0.02 mas and coronal bending at 15 GHz was
0.03 mas, less than the position errors for each day.  Hence, the
positions on these two dates were chosen to be the undeflected
relative positions at each frequency among the sources.  The position
differences between the two days are generally consistent with the
estimated errors, although there are several outliers are about
3-$\sigma$, especially for the N/S positions for J1246 and J1248.

\subsection {Position Changes with Session and Frequency}

Table~4 lists the position changes from the non-deflected positions
for the three sources with respect to J1304, as a function of session
and frequency.  The assumed {\it non-deflected} position is given on
the first line for each source and frequency, and it is the weighted
average of the October 01 and October 18 positions, given in Table 3.
The offsets of about 1 mas for 3C279, J1246 and J1248 reflect the
error in the apriori position of J1304.  The offset positions among
3C279, J1246 and J1248 are in good agreement and suggests that their
apriori positions are consistent to about 0.3 mas.  The slight
differences in the offset positions among the three frequencies are
real and show the affect of sturcture changes for each source among
the frequencies.  A more complete discussion of the registration of
the source positions among the three frequencies is given elselwhere
(Fomalont et al, in preparation).  

Below this line, the table then lists the position changes from the
undeflected positions for the other observing dates.  Since the
gravitational bending with $\gamma=1$ between the sources was included
in the correlator model, the position changes will include only the
gravitational bending associated with a ($\gamma-1$) term, as well as
that caused by noise and remaining tropospheric and coronal refraction
components that differ among the sources.

The first line for an observing date, labeled {\it Each Freq}, lists
the residual positions of the sources for each of the three
frequencies.  All entries are obtained from the relevant entries in
Table~3.  The second line for an observing date, labeled {\it Corona
Free}, gives the residual coronal-free position.  Since the coronal
plasma deflection varies with the observation wavelength squared at
radio frequencies \citep{tho01}, the estimate of the coronal-free
position, ${\bf p_0}$, can be obtained from the relationship,

$${\bf p}(\nu_i) = {\bf p_0} + {\bf I} \nu_i^{-2}$$

\noindent where ${\bf p}(\nu_i)$ are the measured positions (x,y) at
the three frequencies $\nu_i$, and ${\bf I}$ is the average coronal
refraction (magnitude and direction in the sky) at 1 GHz for the
observation period \citep{rog70}.  We determined the estimated values
and uncertainty of the ionosphere correction and the corona-free
position of the sources using a least-square fit.  The discussion of
the effectiveness in obtaining corona-removed solutions is given in
the next section.

\section{The Determination of the Parameter $\gamma$}

The approximate accuracy for the determination of $\gamma$ from Table
4 can be estimated in a straight-forward manner.  Since the typical
position uncertainty of one measurement (one source-pair per day) is
$\sim 0.07$ mas for the differential gravitational bending of $\sim
150$ mas\footnote{For example, On October 11 the average relative deflection
between 3C279 and J1304 was 100 mas and 66 mas in the e/w and n/s directions,
respectively}, the measurement accuracy of 1 part in 2000 corresponds
to an accuracy of $\gamma$ of 1 part in 1000.  Since about 20
independent measurements of the bending were made (10 source pairs
with two position coordinates), a sensitivity of $\gamma$ to a few
parts in $10^{-4}$ is expected.

The more formal determination of the optimum value of ($\gamma-1$)
from the experimental results in Table~4 is straight-forward.  The
analysis minimize the (normalized) chi-squared expression

$$ \chi_k^2 = \frac{1}{k}\sum_{d,i} \Big( \frac {P_d(i) - 0.5(\gamma-1)D_d(i)}{\sigma_d(i)}\Big)^2 $$

\noindent where $P_d(i)$ and $\sigma_d(i)$ are the measured position
offset and error estimate, respectively, from Table 4.  The ($i$)
loops over the sources (3C279, J1246, J1248), and the ($d$) loops over
each day or half-day observation (Oct05, 09, 10a, 10b, 11).  The term
$D_d(i)$ is the differential General Relativity gravitational bending
prediction, averaged over the session.  The sum is made over the E/W
and N/S values and estimated errors separately, so the number of
degrees of freedom is $k=19$, if all ten observing points are used.

The results for several solutions for $\gamma$ using the different
data sets from Table~4 are listed in Table~5.  The normalized $\chi^2$
indicates the ratio of the rms of the best fit divided by that
expected from the estimate error of each entry.  The 43 GHz
corona-free fit has the lowest $\chi^2$ for two reasons: the lessening
of some coronal effects, and the increase of the position errors.  The
two 43-GHz only solutions (with no removal of the ionosphere
contribution) show the effect of the Oct05 session that was made
relatively close to the sun.  Finally, the 23-GHz only solution has a
relatively large normalized $\chi^2$, even excluding the Oct05
session, and suggests that coronal refraction, which is four times
larger than that at 43 GHz, is dominating the sensitivity of the
experiment at 23 GHz.  Nevertheless, the variation of $\gamma$ and its
estimated error among the different solutions are in good agreement
and suggest that the determination of $\gamma$ change little between
several different analyses of the data.

\section{Results and Discussion}

The solutions in Table~5, using different selections of data, are all
consistent with GR value of the parameter $\gamma=1$.  For the results
in this paper, we have taken an average of the four solutions to
obtain $\gamma=0.9998\pm 0.0003$.  This result is a factor of three
times more accurate than the previous {\it dedicated} radio
interferometric observations made specifically to measure the
gravitational bending: 0.9996$\pm$ 0.0017 \citep{leb95}, and
$1.0002\pm 0.0010$ \citep{rob91}.  Using the radio geodetic data base
of thousands of observing sessions between 1979 and 1999, \cite{sha04}
obtained $\gamma=0.9998\pm 0.0004$, comparable in accuracy to the
present results.  The result from the measurements as the spacecraft
Cassini passed by the sun in September 2002 is $\gamma = 1.00002\pm
0.00002$ \citep{ber03}, and was discussed in \S 1.

Although expected departures from $\gamma=1$ are likely to be a factor
of 10 to 100 smaller than the accuracy of this experiment, some
possibilities are: (1) The long-range scalar-tensor interaction in
scalar-tensor theories of gravity \citep{wil93}, and \cite{dam93}
predict a lower bound for the present value of $\gamma$ at the level
of $10^{-6}\sim 10^{-7}$; (2) The long-range vector-tensor interaction
in vector-tensor theories of gravity \citep{kos04} may suggest a
\lq\lq spontaneous violation\rq\rq~of the Lorentz invariance and could
modify the value of $\gamma$ \citep{bai06}; (3) The more complicated
nature of the gravitational coupling between the curvature and
stress-energy tensor of matter may lead to changes of $\gamma$
\citep{jae05}; (4) The plausible existence of the effective graviton's
mass \citep{bab03} that would avoid the van Dam-Veltmann-Zakharov
discontinuity \citep{van70,zakh70} would also effect the value of
$\gamma$.

There are several changes in the design of a similar VLBA experiment
that should improve the accuracy of $\gamma$ by about a factor of two.
First, by choosing a set of sources that can be observed when the sun
is further north\footnote{Several possible experiments are: March 30
to April 24---J0121+1149, J0104+1134, J0139+0842, J0129+1146; April 10
to May 01---J0204+15114, J0158+1307, J0209+1352, J0211+1051}, each
day's integration time can be increased from 6 to 10 hours, and the
sources will be on average at higher elevations than those from the
present experiment.  We estimate that a more northern experiment will
lower the position rms by about 20\%.

Secondly, we lost some sensitivity in the determination of $\gamma$ by
not observing on days when the gravitational deflection was relatively
large (egs October 3, 4, 12, 13).  We found that the most accurate results
are obtained between solar distances of three to five degrees,
corresponding to about three days to seven days from the nearest
approach of the sources with the sun.  Thus, the addition of the above
four observing days in the 2005 experiment would have decreased the
position rms by 25\%.

Finally, the 2005 experiment devoted too much observation time at 15
GHz and possibly 23 GHz in an attempt to lessen the effect of coronal
bending.  When the coronal bending was significant at 43 GHz ($>0.1$
mas), the phase stability degraded considerably at all frequencies and
no astrometric information could be obtained.  Thus, future
observations should concentrate on 43 GHz, somewhat further from the
sun.  Additional observations at 23 GHz are useful to monitor the
coronal refraction since it can be variable with time, but
observations at 15 GHz are not helpful in determining the coronal
refraction.  The potential additional observing time by a factor of
two at the primary frequency of 43 GHz in a new experiment, compared
with the 2005 experiment, would decrease the position rms by about
30\%.

Hence, the above three changes of strategy in a VLBA-designed
experiment will clearly improve the accuracy of the determination of
$\gamma$ from a single 10-day experiment by at least about a factor of
two compared with that of the October 2005 experiment.  The
availability of many such experiments over the year, all repeatable
from year to year, can provide independent results estimates of
$\gamma$ since the dominant error is produced by the quasi-random
errors of the short-term troposphere and, to a lesser extent, coronal
refraction.  Also, any small systematic errors associated with the
different sky configuration for the sources among the other possible
experiments should be somewhat independent.  Thus, we expected that
the accuracy of $\gamma$ should improve by roughly the square-root of
the number of experiments.  Hence, with sufficient VLBA resources, the
uncertainty in the parameter $\gamma$ can be decreased by at least a
factor of four compared with the results given in this paper.

\section {Acknowledgments}

The National Radio Astronomy Observatory is a facility of the National
Science Foundation operated under cooperative agreement by Associated
Universities, Inc.  Research performed at the Jet Propulsion
Laboratory is supported by a NASA contract with the California
Institute of Technology.

\clearpage

\begin{figure} 
\includegraphics{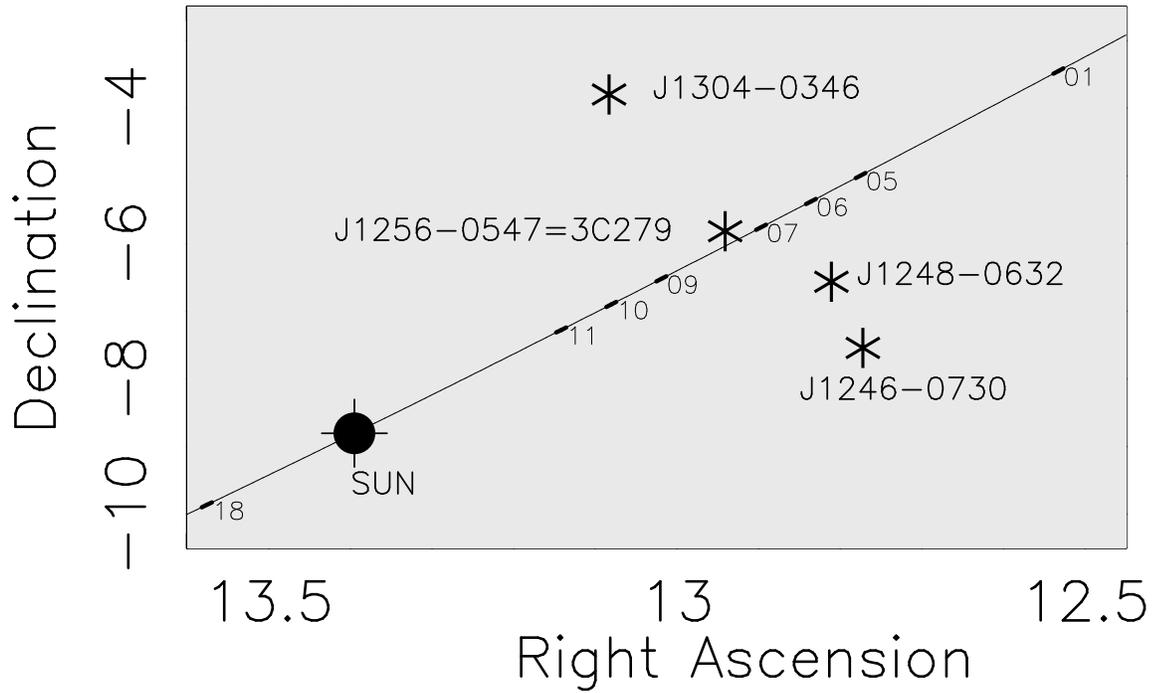} 
\vspace{10cm} 

\caption {{\bf The Source Configuration for the Deflection Experiment:}
The solar trajectory between October 1 and 18 is shown by the diagonal
line, with the eight observing days superimposed.  The location of the
four radio sources are indicated.}

\end{figure}

\clearpage

\begin{figure} 
\includegraphics{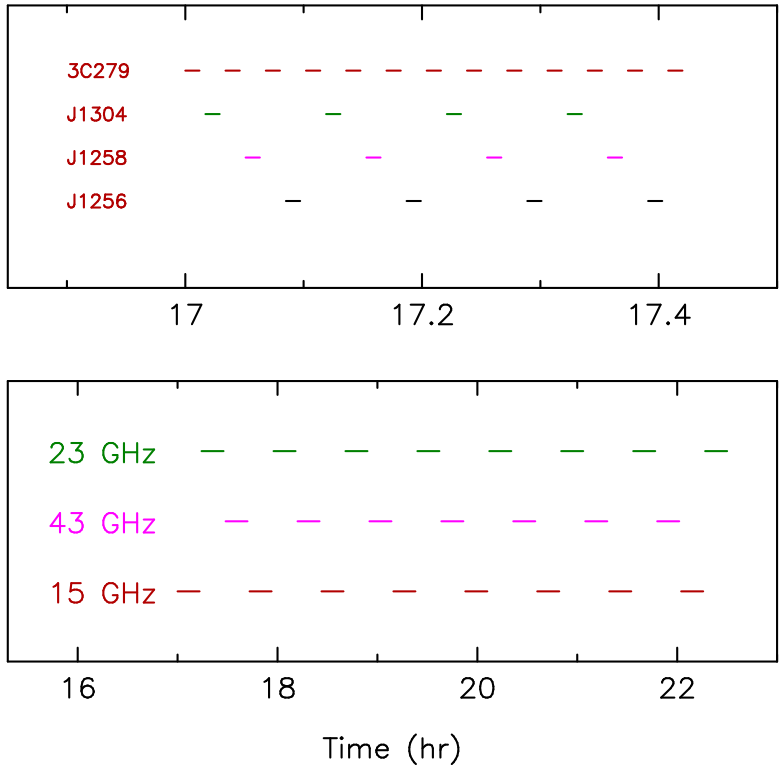} 
\vspace{10cm} 

\caption {{\bf The Observing Strategy:} The top plot shows the
switching among the four sources within every 20-minute period.  The
bottom plot shows the switching of these 20-min blocks among the three
frequencies through each observing day.}

\end{figure}

\clearpage

\begin{figure} 
\includegraphics{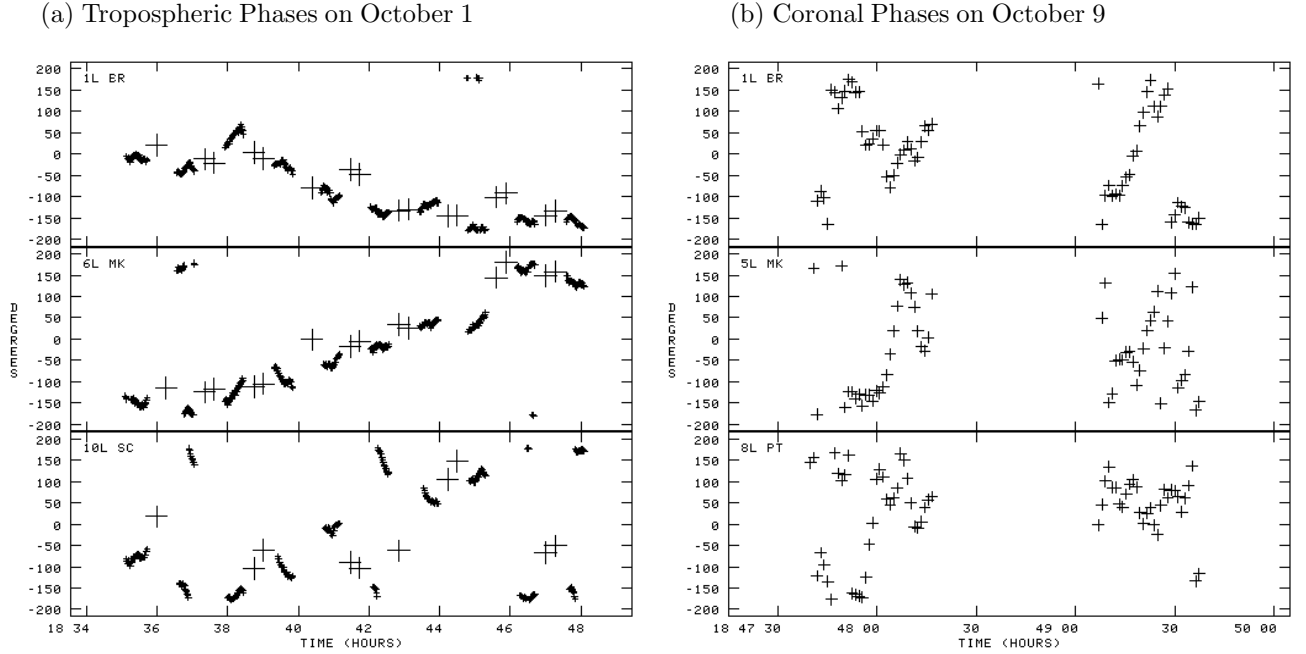} 
\vspace{10cm} 

\caption {{\bf (a) Tropospheric Phase Behavior}: The phase behavior at
23 GHz on October 1, when the sources were $>5^\circ$ from the sun, is
shown for three VLBA baselines from Los Alamos, NM (LA) to: Brewster,
WA (BR), 1800 km; Mauna Kea, HI (MK), 5000 km; and Saint Croix (SC),
VI, 4500 km.  The period covers 13 minutes near the middle of the day,
with the UT time given at the bottom.  The densely-packed points are
the phases for 3C279 at two-second intervals, and the more isolated
larger points show the phases for the other calibrators, averaged over
20 sec (two per scan).  All sources lie on a relatively continuous
temporal phase curve because their accurate relative positions were
determined before producing the plot.  {\bf (b) The Coronal Phase
Behavior:} The phase behavior at 23 GHz on October 9 for 3C279, when
the source was $1.7^\circ$ from the sun, is shown for three VLBA
baselines over a 2-min period.  The baseline between LA and Pie Town,
NM (PT) is only 240 km.  Each plotted point shows the measured phase
for 3C279 at two-second intervals.  The other signal from the other
three sources observed during the gaps were too decorrelated to be
detected.}

\end{figure}

\clearpage

\begin{deluxetable}{lrcrrr}
\tablecolumns{6}
\tablewidth{0pt}
\tabletypesize{\footnotesize}
\tablecaption{Apriori Radio Source Position and Total Flux Density}
\tablehead {
 \colhead {Source} &
 \colhead {RA} &
 \colhead {DEC} &
 \colhead {S$_{43 GHz}$} &
 \colhead {S$_{23 GHz}$} &
 \colhead {S$_{15 GHz}$} \\
 \colhead {}    &
 \multicolumn{2}{c}{Equinox 2000.0} &
 \multicolumn{3}{c}{Jy} \\
}
\startdata

3C279      & 12 56 11.166557$\pm$ 0.000013  & $-$05 47 21.52467$\pm$ 0.00031  & 11.22 & 12.75 & 15.27  \\
J1246-0730 & 12 46 04.232100$\pm$ 0.000014  & $-$07 30 46.57456$\pm$ 0.00031  &  0.24 &  0.34 &  0.50  \\
J1248-0632 & 12 48 22.975657$\pm$ 0.000016  & $-$06 32 09.81737$\pm$ 0.00041  &  0.16 &  0.19 &  0.26  \\
J1304-0346 & 13 04 43.642255$\pm$ 0.000022  & $-$03 46 02.55122$\pm$ 0.00065  &  0.38 &  0.48 &  0.65  \\ 
\enddata
\end{deluxetable}
\clearpage

\begin{deluxetable}{rrrrr}
\tablecolumns{5}
\tablewidth{0pt}
\tabletypesize{\normalsize}
\tablecaption{Peak Flux Density Variations of Sources}
\tablehead {
 \colhead {Date/Freq} &
 \colhead {3C279} &
 \colhead {J1246} &
 \colhead {J1248} &
 \colhead {J1304} \\
}
\startdata
Oct 01:43 GHz & 10.1$\pm$ 0.3 & 0.50$\pm$ 0.02 & 0.40$\pm$ 0.02 & 0.44$\pm$ 0.03 \\
Oct 18:43 GHz &  9.0$\pm$ 0.4 & 0.54$\pm$ 0.02 & 0.48$\pm$ 0.02 & 0.51$\pm$ 0.03 \\
\\
Oct 01:23 GHz & 12.9$\pm$ 0.2 & 0.61$\pm$ 0.02 & 0.41$\pm$ 0.03 & 0.52$\pm$ 0.02 \\
Oct 18:23 GHz & 11.2$\pm$ 0.4 & 0.63$\pm$ 0.02 & 0.50$\pm$ 0.04 & 0.57$\pm$ 0.03 \\
\\
Oct 01:15 GHz & 13.7$\pm$ 0.7 & 0.62$\pm$ 0.02 & 0.28$\pm$ 0.02 & 0.51$\pm$ 0.03 \\
Oct 18:15 GHz & 12.9$\pm$ 0.3 & 0.59$\pm$ 0.02 & 0.27$\pm$ 0.02 & 0.49$\pm$ 0.04 \\

\enddata
\end{deluxetable}

\begin{deluxetable}{lrrrrrrr}
\tablecolumns{8}
\tablewidth{0pt}
\tabletypesize{\scriptsize}
\tablecaption{Source Positions with Respect to J1304}
\tablehead {
 \colhead {Source}  & 
 \colhead {S$_{R\odot}$} &
 \multicolumn{2} {c} {43 GHz} &
 \multicolumn{2} {c} {23 GHz} &
 \multicolumn{2} {c} {15 GHz} \\
 \colhead {Date}  &
 \colhead {(deg)} &
 \colhead {E/W} & \colhead {N/S} &
 \colhead {E/W} & \colhead {N/S} &
 \colhead {E/W} & \colhead {N/S} \\
 }
\startdata
{\bf 3C279}   \\
 Oct01  & 6.6 & $  0.96\pm 0.03 $ & $ -0.11\pm 0.04 $ & $  0.77\pm 0.03 $ & $  0.16\pm 0.04 $ & $  0.78\pm 0.04 $ & $ -0.05\pm 0.05 $   \\
 Oct05  & 2.8  \\
 Oct09  & 1.3  \\
 Oct10a & 2.2 & $  0.95\pm 0.10 $ & $ -0.09\pm 0.14 $ & $  0.65\pm 0.14 $ & $ -1.27\pm 0.20 $ & $  0.36\pm 0.23 $ & $ -2.30\pm 0.30 $   \\
 Oct10b & 2.4 & $  0.93\pm 0.13 $ & $ -0.17\pm 0.18 $ & $  0.56\pm 0.16 $ & $ -1.28\pm 0.24 $ & $  0.20\pm 0.28 $ & $ -2.80\pm 0.40 $   \\
 Oct11  & 3.2 & $  1.03\pm 0.03 $ & $ -0.11\pm 0.04 $ & $  1.02\pm 0.04 $ & $ -0.05\pm 0.05 $ & $  1.11\pm 0.07 $ & $ -0.29\pm 0.09 $   \\
 Oct18  &10.2 & $  0.93\pm 0.02 $ & $ -0.08\pm 0.03 $ & $  0.83\pm 0.02 $ & $  0.16\pm 0.04 $ & $  0.84\pm 0.04 $ & $ -0.02\pm 0.05 $   \\
\tableline
{\bf J1246}   \\
 Oct01 & 5.5 & $  0.75\pm 0.05 $ & $  0.16\pm 0.07 $ & $  0.64\pm 0.05 $ & $  0.54\pm 0.07 $ & $  0.81\pm 0.05 $ & $  0.33\pm 0.07 $   \\
 Oct05 & 2.6 & $  0.46\pm 0.06 $ & $  0.15\pm 0.10 $ & $ -0.23\pm 0.08 $ & $  0.49\pm 0.11 $ & $ -0.22\pm 0.10 $ & $  2.15\pm 0.14 $   \\
 Oct09 & 3.7 & $  0.76\pm 0.05 $ & $  0.09\pm 0.07 $ & $  0.82\pm 0.07 $ & $  0.28\pm 0.11 $ & $  0.78\pm 0.05 $ & $  0.16\pm 0.07 $   \\
 Oct10 & 4.5 & $  0.76\pm 0.05 $ & $  0.09\pm 0.07 $ & $  0.82\pm 0.07 $ & $  0.28\pm 0.11 $ & $  0.78\pm 0.05 $ & $  0.16\pm 0.07 $   \\
 Oct11 & 5.4 & $  0.75\pm 0.08 $ & $ -0.09\pm 0.12 $ & $  0.66\pm 0.10 $ & $  0.29\pm 0.14 $ & $  0.75\pm 0.15 $ & $  0.18\pm 0.10 $   \\
 Oct18 &12.1 & $  0.71\pm 0.05 $ & $  0.18\pm 0.07 $ & $  0.68\pm 0.05 $ & $  0.38\pm 0.07 $ & $  0.86\pm 0.05 $ & $  0.21\pm 0.07 $   \\
\tableline
{\bf J1248}   \\
 Oct01  &5.3 & $  0.70\pm 0.05 $ & $ -0.40\pm 0.07 $ & $  0.66\pm 0.05 $ & $ -0.02\pm 0.07 $ & $  0.85\pm 0.05 $ & $ -0.15\pm 0.07 $   \\
 Oct05  &1.8 \\
 Oct09  &3.0 & $  0.68\pm 0.09 $ & $ -0.32\pm 0.11 $ & $  0.80\pm 0.18 $ & $  0.00\pm 0.23 $ & $  1.02\pm 0.10 $ & $  0.10\pm 0.14 $   \\
 Oct10  &3.9 & $  0.71\pm 0.13 $ & $ -0.13\pm 0.15 $ & $  0.65\pm 0.13 $ & $ -0.19\pm 0.19 $ & $ -0.59\pm 0.21 $ & $  1.19\pm 0.33 $   \\
 Oct11  &4.9 & $  0.70\pm 0.08 $ & $ -0.49\pm 0.12 $ & $  0.74\pm 0.10 $ & $ -0.40\pm 0.15 $ & $  0.64\pm 0.15 $ & $ -0.84\pm 0.20 $   \\
 Oct18 &11.8 & $  0.65\pm 0.05 $ & $ -0.41\pm 0.10 $ & $  0.67\pm 0.05 $ & $ -0.25\pm 0.07 $ & $  0.85\pm 0.05 $ & $ -0.21\pm 0.07 $   \\

\enddata
\end{deluxetable}

\clearpage 

\begin{deluxetable}{lcrrrrrr}
\tablecolumns{8} 
\tablewidth{0pt}
\tabletypesize{\scriptsize}
\tablecaption{Deflection Changes Between the Sources and J1304 }
\tablehead {
 \multicolumn{1}{l} {DATE} &
 \multicolumn{1}{c} {Type} &
 \multicolumn{2}{c} {43 GHz} &
 \multicolumn{2}{c} {23 GHz} &
 \multicolumn{2}{c} {15 GHz} \\

 \colhead { } & \colhead { } &
 \multicolumn{1}{c} {E/W} &
 \multicolumn{1}{c} {N/S} &
 \multicolumn{1}{c} {E/W} &
 \multicolumn{1}{c} {N/S} &
 \multicolumn{1}{c} {E/W} &
 \multicolumn{1}{c} {N/S}  \\

 \colhead { } & \colhead { } &
 \multicolumn{2}{c} {(mas)} &
 \multicolumn{2}{c} {(mas)} &
 \multicolumn{2}{c} {(mas)} \\
}
\startdata
{\bf 3C279} \\
\multicolumn{2}{l} {nondeflected position}& $    0.94~~~~~~~~~ $ & $   -0.09~~~~~~~~~ $ & $    0.80~~~~~~~~~ $ & $    0.16~~~~~~~~~ $ & $    0.81~~~~~~~~~ $ & $   -0.04~~~~~~~~~ $ \\
Oct10a & Each Freq  & $    0.00\pm    0.10 $ & $     0.00\pm    0.14 $     & $    -0.15\pm    0.14 $ & $    -1.43\pm    0.20 $     & $    -0.45\pm    0.23 $ & $    -2.26\pm    0.30 $    \\
Oct10a & Corona-Free& $    0.06\pm    0.11 $ & $     0.39\pm    0.20 $ &&&& \\    
Oct10b & Each Freq  & $   -0.01\pm    0.13 $ & $    -0.08\pm    0.18 $     & $    -0.24\pm    0.16 $ & $    -1.44\pm    0.24 $     & $    -0.61\pm    0.28 $ & $    -2.76\pm    0.40 $    \\
Oct10b & Corona-Free& $    0.06\pm    0.14 $ & $     0.33\pm    0.25 $ &&&& \\    
Oct11  & Each Freq  & $    0.08\pm    0.03 $ & $    -0.02\pm    0.04 $     & $     0.22\pm    0.04 $ & $    -0.21\pm    0.05 $     & $     0.30\pm    0.07 $ & $    -0.25\pm    0.09 $    \\
Oct11  & Corona-Free& $    0.04\pm    0.05 $ & $     0.03\pm    0.04 $ &&&& \\    
\hline
{\bf J1246} \\
\multicolumn{2}{l} {nondeflected position}& $    0.73~~~~~~~~~ $ & $    0.17~~~~~~~~~ $ & $    0.66~~~~~~~~~ $ & $    0.46~~~~~~~~~ $ & $    0.84~~~~~~~~~ $ & $    0.27~~~~~~~~~ $ \\
Oct05  & Each Freq  & $   -0.27\pm    0.06 $ & $    -0.02\pm    0.10 $     & $    -0.89\pm    0.08 $ & $     0.03\pm    0.11 $     & $    -1.06\pm    0.10 $ & $     1.88\pm    0.14 $    \\
Oct05  & Corona-Free& $   -0.10\pm    0.15 $ & $    -0.11\pm    0.15 $ &&&& \\    
Oct09  & Each Freq  & $    0.03\pm    0.05 $ & $    -0.08\pm    0.07 $     & $     0.16\pm    0.07 $ & $    -0.18\pm    0.11 $     & $    -0.06\pm    0.05 $ & $    -0.11\pm    0.07 $    \\
Oct09  & Corona-Free& $    0.02\pm    0.06 $ & $    -0.05\pm    0.07 $ &&&& \\    
Oct10  & Each Freq  & $    0.00\pm    0.10 $ & $     0.19\pm    0.13 $     & $     0.06\pm    0.13 $ & $    -0.41\pm    0.18 $     & $     0.11\pm    0.18 $ & $    -0.34\pm    0.31 $    \\
Oct10  & Corona-Free& $   -0.02\pm    0.11 $ & $     0.28\pm    0.14 $ &&&& \\    
Oct11  & Each Freq  & $    0.02\pm    0.08 $ & $    -0.26\pm    0.12 $     & $     0.00\pm    0.10 $ & $    -0.17\pm    0.14 $     & $    -0.09\pm    0.15 $ & $    -0.09\pm    0.10 $    \\
Oct11  & Corona-Free& $    0.03\pm    0.09 $ & $    -0.23\pm    0.12 $ &&&& \\    
\hline
{\bf J1248} \\
\multicolumn{2}{l}{nondeflected position}& $    0.67~~~~~~~~~ $ & $   -0.41~~~~~~~~~ $ & $    0.67~~~~~~~~~ $ & $   -0.14~~~~~~~~~ $ & $    0.85~~~~~~~~~ $ & $   -0.18~~~~~~~~~ $ \\
Oct09  & Each Freq  & $    0.01\pm    0.09 $ & $     0.09\pm    0.11 $     & $     0.13\pm    0.18 $ & $     0.14\pm    0.23 $     & $     0.17\pm    0.10 $ & $     0.28\pm    0.14 $    \\
Oct09  & Corona-Free& $   -0.02\pm    0.11 $ & $     0.05\pm    0.11 $ &&&& \\    
Oct10  & Each Freq  & $    0.04\pm    0.13 $ & $     0.28\pm    0.15 $     & $    -0.02\pm    0.13 $ & $    -0.05\pm    0.19 $     & $    -1.44\pm    0.21 $ & $     1.37\pm    0.33 $    \\
Oct10  & Corona-Free& $    0.09\pm    0.16 $ & $     0.20\pm    0.18 $ &&&& \\    
Oct11  & Each Freq  & $    0.03\pm    0.08 $ & $    -0.09\pm    0.12 $     & $     0.07\pm    0.10 $ & $    -0.26\pm    0.15 $     & $    -0.21\pm    0.15 $ & $    -0.66\pm    0.20 $    \\
Oct11  & Corona-Free& $    0.03\pm    0.09 $ & $    -0.01\pm    0.13 $ &&&& \\    
\hline

\enddata
\end{deluxetable}
\clearpage

\begin{deluxetable}{lccc}
\tablecolumns{4}
\tablewidth{0pt}
\tabletypesize{\normalsize}
\tablecaption{Solutions for $\gamma$}
\tablehead {
 \colhead {Solution Type} &
 \colhead {$\gamma-1$} &
 \colhead {$\sigma_{\gamma}$} &
 \colhead {$\chi_k^2$} \\
 \colhead {} &
 \colhead {$10^{-4}$} &
 \colhead {$10^{-4}$} \\
}
\startdata

43 GHz data (corona-free)   &  -2.4  &  3.2  &  0.9  \\
43 GHz data only            &  -1.0  &  2.6  &  2.2  \\
43 GHz data only - Oct5a    &  -3.2  &  2.8  &  1.1  \\ 
23 GHz data only - Oct5a    &  -2.0  &  2.4  &  4.7  \\
\enddata
\end{deluxetable}

\end{document}